\documentclass{iopjournal}

\usepackage{amsmath}
\usepackage{amsfonts}
\usepackage{amssymb}
\usepackage{amsthm}
\usepackage{bbm}
\usepackage{mathtools}
\usepackage{mathabx}
\usepackage{comment}
\usepackage{color}
\usepackage{mathrsfs}
\usepackage{hyperref}

\DeclareMathOperator{\tr}{tr}
\DeclareMathOperator{\Tr}{Tr}

\begin{document}

\title{Entanglement Entropy in Loop Quantum Gravity through Quantum Error Correction}

\author{Sean Tobin$^1$}

\affil{$^1$Department of Physics, Florida Atlantic University, 777 Glades Road, Boca Raton, FL 33431-0991, USA}

\email{stobin2023@fau.edu}

\begin{abstract}
We introduce a novel method for computing entanglement entropy across surfaces in Loop Quantum Gravity by employing techniques from quantum error correcting codes. In this construction, the redundancy encoded in the gauge invariant subspace is made manifest by embedding it in a larger Hilbert space. The enlarged Hilbert space of a surface does not factorize, which necessitates an algebraic formulation of the entanglement entropy using von Neumann algebras. Using this approach, we are able to reproduce the expected black hole entropy through the canonical ensemble. This includes a direct realization of the Ryu-Takayanagi formula, providing a first principles derivation of the black hole entropy within a kinematical framework of loop quantum gravity. The algebraic techniques developed in this work can be used to compute the entanglement entropy across arbitrary surfaces.
\end{abstract}

\section{Introduction}
Understanding black hole entropy remains a persistent puzzle at the core of quantum gravity. It has been theoretically established that quantum effects drastically change the nature of black holes. Semiclassical analyses have shown that generic black holes radiate as perfect black bodies with an entropy proportional to the horizon area  \cite{hawking1975particlecreation}. Developments made in AdS/CFT and holography have made connections between the entropy of a horizon and quantum information \cite{ryu2006holographicentropy}. These developments pave the way to apply information theoretic techniques to Loop Quantum Gravity (LQG) in order to derive the semiclassical results from first principles. 

In this work, we present a novel approach to compute the entanglement entropy across surfaces within LQG by employing techniques first developed in quantum information theory: quantum error correcting codes. Quantum error correcting codes benefit from embedding states in larger Hilbert spaces so that the original code can be recovered from an erasure due to the additional redundancy. In the context of entanglement across a quantum surface, the idea is similar. At the kinematical level, the geometry must be $SU(2)$ gauge invariant \cite{ashtekar2004status}. The redundancy from gauge invariance can be exploited when we embed the invariant states into a larger non-gauge invariant Hilbert space. Erasures naturally appear in the black hole context; an observer outside of the event horizon only has access to a reduced collection of the degrees of freedom. The entanglement entropy fundamentally relies on the interplay between the gauge invariance and reduced density matrices. 

To deal with the reduced degrees of freedom, reduced density matrices become central to the story. If the Hilbert space factorizes, then the reduced states are trivial to construct. When the Hilbert space factorizes, the partial trace over a factor exists. Due to the direct sum structure inherent to LQG Hilbert spaces, more general techniques are needed than the conventional partial traces over regions \cite{sorce2023notes}. This purely algebraic formalism is based on von Neumann algebras and allows us to rigorously define an entanglement entropy across any closed surface. Critically, we are able to derive the Bekenstein-Hawking entropy by extremizing the entropy subject to an area constraint, analogous to the canonical ensemble. Previous ensemble approaches suffer from introducing the number of surface punctures from the spin-network state as a macroscopic variable \cite{ghosh2011isolatedhorizons}. Our algebra-first approach circumvents this by writing the Hilbert space as a sum over punctures and spin configurations. Furthermore, our formalism naturally reproduces the Ryu-Takayanagi (RT) formula. Within the context of holography, this formula establishes a direct link between entanglement entropy and geometric degrees of freedom encoded on the boundary \cite{harlow2016rtqec}.

Our methodology relies on mapping the geometric states to their quantum error correction counterparts \cite{knill1997conditions}. The LQG Hilbert space of gauge invariant states plays the role of the code subspace. The code subspace is embedded into a larger auxiliary space in analogy with the encoding of logical states. Through this encoding, it is clear to see how the gauge invariance enforces the entanglement across the surface. The resulting Hilbert space does not factorize. The use of von Neumann algebras allows us to construct the reduced density matrices, without invoking a partial trace, by utilizing the algebraic decomposition of the Hilbert space. Additionally, the embedding of the gauge invariant states into the larger Hilbert space removes the potential ambiguity in defining a trace on the von Neumann algebra. The traces on the algebra are only unique up to scaling.

With a general construction for the entropy of a state in hand, the entropy across any surface can be computed by specifying the microscopic degrees of freedom. Treating the state as a statistical ensemble, we are able to derive the distribution of configurations by extremizing the entropy while keeping the average area of the surface fixed. The resulting distribution maximizes the entropy, hence why the state is interpreted as representing a black hole. The resulting entropy satisfies the semiclassical area law form but with the added generality that the calculation did not require any semiclassical input:
\begin{equation}
    S(\rho,\mathcal{A}) = \frac{\lambda}{\gamma}\frac{A}{4\ell_p^2}+\ln \mathcal{Z}
\end{equation}
As a result, extra care is needed in order to make contact with the semiclassical analyses and previous computations within LQG. In particular, there is the issue of renormalization of the couplings and fixing the value of the Lagrange multiplier enforcing the area constraint. However, our construction is not fully general; it is only kinematical. The impact of the dynamics buried within the diffeomorphism and Hamiltonian constraints have been left for future research. Nonetheless, the anticipated behavior of the entanglement entropy can already be seen.

This paper is structured as follows. In section \ref{sec:QEC}, we review the essential concepts from quantum error correcting codes in order to highlight how the redundancy leads to the entanglement. In section \ref{sec:Single_Edge}, we illustrate how this entanglement manifests from the gauge invariance within LQG by considering a surface punctured by a single edge. This toy model contains many features of the full Hilbert space, but the enlarged Hilbert space factorizes. In section \ref{sec:Algebra}, we develop the mathematical machinery needed to handle the nonfactor Hilbert spaces. The von Neumann algebras and renormalized trace will allow us to construct reduced density matrices and their entropy. In section \ref{sec:Entropy_Law}, we develop our main results by considering the Hilbert space of a surface punctured by general spin-network states. We explicitly derive the black hole entropy and conclude by comparing our result with existing computations in the literature.

\section{Quantum Error Correction}\label{sec:QEC}
Quantum error correction (QEC) provides a robust framework for protecting quantum information through entanglement. As an example, consider the qutrit code \cite{cleve1999quantumsecret, harlow2016rtqec}. Consider a normalized vector in the single qutrit Hilbert space $H_{\ell}$:
\begin{equation}
    |\phi\rangle = \alpha|0\rangle+\beta|1\rangle+\gamma|2\rangle
\end{equation}
For example, if Alice sends a quantum state $\rho = |\phi\rangle\langle\phi|$ to Bob through a noisy channel, the state may be corrupted. In the event of an erasure, the information in a portion of the Hilbert space is lost. Instead of receiving the full state, Bob only receives a reduced state. In principle, Bob would be able to purify the state to recover the full state, but this purification is not unique. 

To overcome the non-uniqueness of purification, Alice can introduce redundancy by embedding the logical Hilbert space $H_{\ell}$ into a larger physical\footnote{This Hilbert space should not be confused with the physical Hilbert space of LQG, where the constraints are satisfied. In this context, physical refers to the actual states that are being sent.} Hilbert space $H_p$. The logical Hilbert space $H_{\ell}$ is embedded as the code subspace $H_{\text{code}}\subset H_p$. For our qutrit code, this can be readily done by using three copies of the logical Hilbert space to make our physical Hilbert space, $H_p = H_{\ell}\otimes H_{\ell}\otimes H_{\ell}$, and mapping the basis of $H_{\ell}$ into the basis of $H_{\text{code}}$:
\begin{align}
    |\widetilde{0}\rangle = \frac{1}{\sqrt{3}}\bigl( |000\rangle+|111\rangle+|222\rangle\bigr)\\
    |\widetilde{1}\rangle = \frac{1}{\sqrt{3}}\bigl( |012\rangle+|120\rangle+|201\rangle\bigr)\\
    |\widetilde{2}\rangle = \frac{1}{\sqrt{3}}\bigl( |021\rangle+|102\rangle+|210\rangle\bigr)
\end{align}
The tilde \;$\widetilde{\phantom{a}}$\; denotes vectors or operators that stem from the code subspace and are embedded into the larger space. The extra redundancy in the physical Hilbert space is what allows for the recovery of the logical qutrit after an erasure.

This embedding fundamentally changes the nature of the state. While $\rho = |\phi\rangle\langle\phi|$ is a pure state on the logical Hilbert space, the corresponding density matrix $\widetilde{\rho}=|\widetilde{\phi}\rangle\langle\widetilde{\phi}|$ is a mixed state with non-zero entropy on the larger physical Hilbert space. We are now sending a state on the code subspace with a nontrivial entanglement which we can exploit to recover the logical state after an erasure.

We can be more explicit about this recovery process. On the code subspace, there exists a unitary map $U_{12}$, which only acts on the first two basis vectors, which allows us to extract the logical qutrit from the physical:
\begin{equation}
    U^{\dagger}_{12}|\widetilde{i}\rangle = |i\rangle_1\otimes |\chi\rangle_{23}
\end{equation}
Here, $|\widetilde{i}\rangle$ is the $i$'th basis vector on $H_p$ and $|\chi\rangle_{23}$ is given by:
\begin{equation}
    |\chi\rangle_{23} = \frac{1}{\sqrt{3}}\bigl(|00\rangle+|11\rangle+|22\rangle\bigr)
\end{equation}
The subscripts denote which basis vectors are being considered. The unitary operator is a permutation map described in \cite{harlow2014bulk}. This unitary map can readily be lifted to its action on states:
\begin{equation}
    \widetilde{\rho}= U_{12}\bigl(\rho_1\otimes |\chi\rangle\langle\chi|_{23}\bigr)U^{\dagger}_{12}
\end{equation}
It is straightforward to see that $\widetilde{\rho}$ is a mixed state on the physical Hilbert space owing to the presence of $|\chi\rangle\langle \chi|$. 

Now, suppose that Alice encodes her logical state and sends it to Bob. If the channel erases the third qutrit, Bob receives the reduced state $\widetilde{\rho}_{12}=\Tr_{3}(\widetilde{\rho})$. The encoded density matrix $\widetilde{\rho}$ yields a predictable result:
\begin{equation}
    \widetilde{\rho}_{12} = U_{12}\bigl( \rho_1\otimes\frac{\mathbbm{1}_2}{3}\bigr)U^{\dagger}_{12}
    \label{eq:QEC_Reduced}
\end{equation}
Bob can act on his received density matrix with $U_{12}$ and easily recover the logical state. Contrast this with what happens in the unencoded case. When Bob receives the reduced state $\rho_{12}$, he must purify it to get the intended pure state. The purification leads to introducing a state $\rho_{3}$ such that $\rho = \rho_{12}\otimes\rho_3$ which acts on the Hilbert space $H = H_{12}\otimes H_3$. However, there is no canonical partial trace $H_3$. The trace is only unique up to an overall scale. Therefore, if Bob only has information about $H_{12}$, then there is no intrinsic way to fix the normalization of the identity on $H_3$. This ambiguity is fixed by the embedding of $H$ as the code subspace into the physical Hilbert space.

The encoding fixes the state $|\chi\rangle\langle \chi|$ and precisely captures the normalization of the identity after the partial trace is taken in the physical Hilbert space. Different choices of physical Hilbert spaces will lead to different normalizations of the trace. Using \eqref{eq:QEC_Reduced} and the induced trace, we find the entropy of the reduced density matrix:
\begin{equation}
    S(\widetilde{\rho}_{12}) = \ln 3+S(\rho_1)
\end{equation}
This result is analogous to the structure of the Ryu-Takayanagi (RT) formula, $S_A = S_{\text{bulk}} + S_{\text{bdry}}$. Here, $S(\rho_1)$ represents the bulk entropy, while the $\ln 3$ term is an analogue of the boundary contribution, arising directly from the entanglement structure of the embedding. We will return to the geometric interpretation and origin of these boundary modes later.

Our goal is to take these tools and recast a state in LQG in terms of quantum error correcting codes. This new language gives a method to understand the entanglement entropy across surfaces. It is also natural to consider redundancy within LQG since there is a kinematical $SU(2)$ gauge symmetry at each vertex and dynamical diffeomorphism invariance. There are two relevant insights that will be instrumental in the future: the need for an enlarged Hilbert space and the need for reduced density matrices. Suppose that we have a region of spacetime separated into two regions, $A$ and $\bar{A}$. An observer in $A$ only has access to reduced density matrices since the degrees of freedom in $\bar{A}$ are inaccessible. Typically, the reduced state $\rho_A$ is related to full state $\rho$ by a partial trace. However, the partial traces only exist when the enlarged Hilbert space factorizes, $H = H_A\otimes H_{\bar{A}}$. For full LQG, the Hilbert space of a surface does not factorize over the regions. To circumvent the non-existence of the partial traces, we need to construct a general state using the machinery of von Neumann algebras in order to find the reduced states and compute their entanglement entropies.

\section{Entanglement Across a Single Edge}\label{sec:Single_Edge}
Before developing the full formalism of von Neumann algebras, we first illustrate how the principles of QEC manifest in a simple LQG toy model. To see this, consider calculating the entanglement entropy across a surface punctured by a single edge. 

The Hilbert space for a single edge is given by $\mathcal{H}_e = L^2(SU(2))$. This space can be decomposed into a direct sum of irreducible representations via the Peter-Weyl theorem:
\begin{equation}
    \mathcal{H}_e = \bigoplus_j V_j\otimes V_{j}^*
\end{equation}
The representation spaces are labeled by half-integers $j$. When the edge punctures a surface, we assume that the surface does not intersect a vertex. As a result, the surface splits the edge into two regions, $A$ and $\bar{A}$, with one vertex residing in each. Since the surface essentially splits the edge in two, we consider the enlarged Hilbert space to be the tensor product of two edges:
\begin{equation}
    \mathcal{H} = L^2_A(SU(2))\otimes L^2_{\bar{A}}(SU(2))
\end{equation}

Drawing a direct parallel to the QEC example, this enlarged space $\mathcal{H}$ serves as the physical Hilbert space. The code subspace, $\mathcal{H}_{\text{code}}$, is then identified with the gauge-invariant subspace of $\mathcal{H}$, which is isomorphic to the original single-edge Hilbert space, $\mathcal{H}_e$. The gauge invariant objects are the $SU(2)$-intertwiners, which are represented by the Wigner $3j$-symbols. If we consider a general vector in $\mathcal{H}$:
\begin{equation}
    |\Psi\rangle = \sum_{j,mn}\sum_{k,op}C^j_{mn}D^k_{op}|j,mn\rangle_A\otimes|k,op\rangle_{\bar{A}}
\end{equation}
The $SU(2)$ intertwiner acts as a gauge invariant projection, yielding vectors in the gauge invariant subspace, $|\widetilde{\Psi}\rangle\in \mathcal{H}_{\text{code}}$:
\begin{equation}
    |\widetilde{\Psi}\rangle = \sum_{j,mn}\sum_{\alpha}C^j_{mn}D^j_{mn}\frac{(-1)^{2j}}{\sqrt{2j+1}}|j,m\alpha\rangle_A\otimes|j,\alpha n\rangle_{\bar{A}}
\end{equation}
Following again from the qutrit example, we can identify the basis vectors on the code subspace:
\begin{equation}
    |\widetilde{j,mn}\rangle = |jm\rangle_A \langle jn|_{\bar{A}}\otimes |\chi_j\rangle
\end{equation}
\begin{equation}
    |\chi_j\rangle = \sum_{\alpha}\frac{(-1)^{2j}}{\sqrt{2j+1}}|j\alpha\rangle_{\bar{A}}\langle j\alpha|_{A}
\end{equation}
The unitary operator is trivial in this case. This embedding of the single-edge Hilbert space into the two-edge Hilbert space can be physically understood as a graph refinement, where the original edge becomes two edges connected with a bivalent vertex. This perspective frames the QEC construction in terms of cylindrical consistency, which ensures that the physics remains consistent across different graphs. Formally, states on a graph $\Gamma$ are mapped to states on a refined graph $\Gamma'$ via an embedding $\iota_{\Gamma\rightarrow\Gamma'}$. The gauge invariant projection exactly enforces the cylindrical consistency between states on a single-edge and two-edge graph. As a result, the QEC structure of embedding a logical Hilbert space into a physical Hilbert space maps directly onto cylindrical consistency in LQG. 

This identification of basis vectors conveniently allows us to express arbitrary vectors in the code subspace in terms of unit vectors and a normalization:
\begin{equation}
    |\widetilde{\Psi}\rangle = \sum_j \sqrt{p_j}|\psi_j\rangle\otimes|\chi_j\rangle
\end{equation}
For a unit vector, the normalization condition is $\sum_j |p_j| = 1$. The unit vector $|\psi_j\rangle$ lives on a mixture of the representation spaces, $|\psi_j\rangle \in V_{j,A}\otimes V^*_{j,\bar{A}}$, while the fixed $|\chi_j\rangle$ lives on the other two pieces, $|\chi_j\rangle \in V_{j,\bar{A}}\otimes V^*_{j,A}$. This $|\chi_j\rangle$ leads to the Ryu-Takayanagi term in the entropy. It captures the entanglement between $A$ and $\bar{A}$. The vector itself is unique up to a unitary transformation and is fixed by embedding the of the gauge invariant Hilbert space into the non-gauge invariant Hilbert space. To see its role more clearly, we compute the entropy of a reduced density matrix. 

If we consider a unit vector $|\widetilde{\Psi}\rangle$ in $\mathcal{H}_{\text{code}}$, we can construct a `pure' density matrix:
\begin{equation}
    \widetilde{\rho} = |\widetilde{\Psi}\rangle\langle \widetilde{\Psi}| = \sum_{jk}\sqrt{p_jp^*_k}|\psi_j\rangle\langle \psi_k|\otimes |\chi_j\rangle\langle \chi_k|
\end{equation}
While this density matrix is `pure' in the sense that $\widetilde{\rho} = |\widetilde{\Psi}\rangle\langle \widetilde{\Psi}|$, it is not `pure' in the sense that it has zero entropy. We saw in the qutrit example that pure states were mapped to mixed states, but this circumstance is different. The classification of density matrices depends on the structure of the operator algebra that $\widetilde{\rho}$ belongs to. In nonrelativistic quantum mechanics, this is hardly an issue because the Hilbert spaces typically have minimal structure and density matrices of the form $|\phi\rangle\langle\phi|$ are the zero entropy states. Our state $\widetilde{\rho}$ is an operator on a Hilbert space that is both infinite dimensional and has a direct sum structure. The $|\phi\rangle\langle\phi|$ `pure' states are no longer the zero entropy states. In fact, pure states do not exist. To better understand this distinction, we need the machinery of von Neumann algebras, which we tackle in the next section.

For now, let's continue with our toy model. Since the enlarged Hilbert space $\mathcal{H}$ has a tensor product structure, we have access to the partial traces. In the qutrit example, this was important since the partial trace is not unique on the code subspace. However, since the gauge invariant subspace does not factorize across $A$ and $\bar{A}$, the partial traces do not exist \cite{lin2018definingentropy}. $L^2(SU(2))$ has a natural trace:
\begin{equation}
    \Tr = \bigoplus_j\tr_j
\end{equation}
Which allows us to determine the reduced density matrix by tracing out $\bar{A}$:
\begin{equation}
    \widetilde{\rho}_A = \sum_{j}p_j\rho_{j,A}\otimes\frac{\mathbbm{1}_j}{2j+1}
\end{equation}
Here $p_j$ is the normalization of the state, which can be interpreted as a probability distribution and $\rho_{j,A} = \tr_{\bar{A}}(|\psi_j\rangle\langle \psi_j|)$. Using the same trace on $L^2_A(SU(2))$, we can compute the entropy of the reduced density matrix:
\begin{equation}
    S(\widetilde{\rho}_A) = -\Tr(\widetilde{\rho}_A\ln\widetilde{\rho}_A) = \sum_j -p_j\ln p_j+p_jS(\rho_{j,A})+\sum_jp_j\ln(2j+1)
\end{equation}
We explicitly see the Ryu-Takayanagi term, $\sum_j p_j\ln(2j+1)$. In the qutrit example, the RT term was $\ln 3$. Here, the RT term stems from the edge mode contributions on the boundary. More explicitly, the form of $\widetilde{\rho}_A$ clarifies that there are two contributions to the state: the bulk and the boundary. The boundary contribution $\frac{\mathbbm{1}_j}{2j+1}$ is the edge mode that appears due to the gauge invariance being enforced at the surface. The edge modes lead to the RT term in the entropy, which reflects that the RT term encodes the geometric degrees of freedom entangled across the boundary. 

Despite the code subspace not having a unique partial trace, we were able to find the reduced density matrices using the partial trace from the enlarged Hilbert space. To handle these more general Hilbert spaces and to clarify the issue about the pure states, we need to first understand von Neumann algebras.

\section{The Basic Theory of von Neumann Algebras}\label{sec:Algebra}
To simplify our discussion, we impose a maximum spin cutoff. Physically, this would correspond to a cosmological constant. This simplification allows us to treat each sector as a finite dimensional Hilbert space. We adapt the more general and physically motivated introduction in \cite{harlow2016rtqec,sorce2023notes} to our direct sum special case.

Given a Hilbert space $H$ and the space of all bounded linear operators acting on the Hilbert space, $\mathcal{L}(H)$, a von Neumann algebra $\mathcal{A}\subset \mathcal{L}(H)$ is a subalgebra of operators that is closed under adjoints and contains the identity. A von Neumann algebra may have a commutant, denoted by $\mathcal{A}'$, which contains all operators in $\mathcal{L}(H)$ that commute with every operator in $\mathcal{A}$. The center of the algebra, defined as the intersection $\mathcal{Z}=\mathcal{A}\cap\mathcal{A}'$, contains all operators that commute with every element of $\mathcal{A}$. The structure of this center is crucial, as it dictates the decomposition of the algebra into fundamental building blocks known as factors. A factor is a von Neumann algebra whose center only contains scalar multiples of the identity operator. 

Factors are important because it can be shown that any von Neumann algebra on a separable Hilbert space is isomorphic to a direct integral of factors. There are three types of factors: type $I$, type $II$ and type $III$. Mathematically, the types are differentiated by the existence of projectors in the algebra. Physically, these projectors correspond to states. Type $I$ algebras contain pure density matrices and mixed density matrices. Type $II$ algebras contain only mixed density matrices. Type $III$ algebras do not contain any density matrices, but do contain states with infinite or zero trace.

This definition is not the full story. Suppose that we have a region of spacetime with a quantum system in it. Are there density matrices in the algebra of observables? Yes, the density matrices are the positive and Hermitian operators with unit trace. Likewise, it is straightforward to construct these states with any orthonormal basis on the Hilbert space. However, what happens if we extend our spacetime to include two causally disconnected regions $A$ and $B$? An observer in region $A$ only has access to the operators $\mathcal{L}(H_A)\otimes \mathbbm{1}_{H_B}$. Does this algebra contain any states? The answer depends on the dimensionality of $H_B$. If $H_B$ is finite dimensional, then we have the state:
\begin{equation}
    \rho\otimes\frac{\mathbbm{1}_{H_B}}{\dim(H_B)}
\end{equation}
However, if $H_B$ is infinite dimensional, then the trace of the identity operator would be infinite and there would be no states. This type of situation readily occurs in quantum field theory.

In this circumstance, we are forced to say that there are no density matrices. However, this is a problem. The resolution stems from considering effective states. Using a renormalized trace, we can consistently assign expectation values to operators $O\otimes \mathbbm{1}_{H_B}$ such that $\Tr_{H_A}(\rho O)$ has all of the expected properties. This ensures that observables are well defined on the region despite the infinities that may appear in the states. Physically, this procedure is analogous to subtraction schemes in quantum field theory. It regularizes the states by factoring out the infinite trace of the identity in the inaccessible region, leaving a well-defined expectation value behind. The type classification refers to the existence of pure and mixed density matrices after a renormalization scheme.

Let's recast our single edge density matrices in the language of von Neumann algebras. The states $\rho$ acting on $L^2(SU(2))$ reside in an algebra of the form $\mathcal{M} \cong \bigoplus_j \mathcal{L}(V_j)\otimes \mathcal{L}(V^*_j)$. Any von Neumann algebra $\mathcal{A}\subset \mathcal{M}$ cannot be a factor since the center would always contain operators of the form:
\begin{equation}
    O_{\mathcal{Z}} = \bigoplus_j \lambda_j(\mathbbm{1}_j\otimes\mathbbm{1}_j)
\end{equation}
Where $\lambda_j$ are arbitrary coefficients as long as $O_{\mathcal{Z}}$ is a state. While the form of our algebra on the code subspace is convenient and simple, it is also quite general. For any von Neumann algebra $\mathcal{A}$, which need not be a factor, on a Hilbert space $\hat{\mathcal{H}}$ then there exists a block diagonal decomposition of the Hilbert space:
\begin{equation}
    \hat{\mathcal{H}} = \bigoplus_{\alpha}\hat{\mathcal{H}}_{A_{\alpha}}\otimes \hat{\mathcal{H}}_{\bar{A}_{\alpha}}
    \label{eq:Factorize}
\end{equation}
such that $\mathcal{A} = \bigoplus_{\alpha}\bigl( \mathcal{L}(\hat{\mathcal{H}}_{A_{\alpha}})\otimes \mathbbm{1}_{\bar{A}_{\alpha}}\bigr)$ and $\mathcal{A}' = \bigoplus_{\alpha}\bigl( \mathbbm{1}_{A_{\alpha}}\otimes \mathcal{L}(\hat{\mathcal{H}}_{\bar{A}_{\alpha}})\bigr)$. Furthermore, for any state $\rho\in\mathcal{M}$ and von Neumann algebra $\mathcal{A}\subset \mathcal{M}$ on $\hat{\mathcal{H}}$, there exists a unique $\rho_{\mathcal{A}}\in \mathcal{A}$ such that $\Tr(\rho O) = \Tr(\rho_{\mathcal{A}}O)$ for all $O\in\mathcal{A}$. The proof of these statements and further discussion can be found in \cite{harlow2016rtqec}. The renormalized trace can also be integrated into these results since it leaves the expectation values invariant \cite{sorce2023notes}. As a result, for any operator $O\in\mathcal{A}$ and state $\rho_{\mathcal{A}}\in\mathcal{A}$:
\begin{equation}
    \Tr(\rho_{\mathcal{A}}O) = \hat{\Tr}(\hat{\rho}_{\mathcal{A}}O)
\end{equation}
This allows us to replace any state $\rho\in\mathcal{M}$ with a state in the von Neumann algebra $\rho_{\mathcal{A}}\in\mathcal{A}$ which in turn can be replaced with the renormalized state and trace:
\begin{equation}
    \Tr(\rho O) = \Tr(\rho_{\mathcal{A}}O) = \hat{\Tr}(\hat{\rho}_{\mathcal{A}}O)
\end{equation}

\subsection{Entropy from a Von Neumann Algebra}
When we discussed quantum error correcting codes, one of the key insights was that Bob could only measure reduced density matrices and still recover the full information in the state. This is the same type of analogy that we want to bring into a gravity perspective. Observers on a spacetime only have access to reduced states. Suppose that there is an observer in a spacetime split into two regions, $A$ and $\bar{A}$. In the simplest scenario the algebra of observables factorizes, $\mathcal{O} = \mathcal{O}_A\otimes \mathcal{O}_{\bar{A}}$. The observer only has access to the subalgebra $\mathcal{O}_{A}\otimes\mathbbm{1}_{\bar{A}}$ which is related to the full algebra by a partial trace.

Without having any reduced density matrices, there is no hope at creating an algebra of observables accessible to an observer. As established in the single-edge model, the direct sum structure of the gauge-invariant Hilbert space prevents the use of a standard partial trace. While the enlarged Hilbert space in that toy model did factorize, this is not true for the general multi-puncture case. The algebraic framework resolves this impasse. In the next section, when we generalize to surfaces punctured by multiple edges, the enlarged Hilbert space does not factorize. However, \eqref{eq:Factorize} comes to the rescue. The von Neumann algebra and its commutant take block diagonal forms that contain states which act as our reduced states. All we need to do is construct them. 

Using the block decomposition theorem, we can identify our von Neumann algebra as the algebra of accessible reduced density matrices. This allows us to define a consistent reduced state without a partial trace. Since we ultimately only care about expectation values, we can restrict our attention to the block diagonal elements, $\rho\sim \bigoplus_{\alpha}\rho_{\alpha\alpha}$. In light of the absent partial trace on the whole space, we still have it on each sector:
\begin{equation}
    \tr_{\bar{A}_{\alpha}}\rho_{\alpha\alpha} = p_{\alpha}\rho_{A_{\alpha}}
\end{equation}
Here $p_{\alpha}$ is a normalization constant that comes from the full state. This defines our reduced density matrix:
\begin{equation}
    \rho_A = \bigoplus_{\alpha}\bigl( p_{\alpha}\rho_{A_{\alpha}}\otimes \frac{\mathbbm{1}_{\partial A_{\alpha}}}{\dim(\hat{\mathcal{H}}^{\text{edge}}_{\alpha})}\bigr)
    \label{eq:Reduced_State}
\end{equation}
Using the trace $\Tr = \bigoplus_{\alpha}\tr_{\alpha}$ on the von Neumann algebra, it is easy to check that $\Tr\rho_A = 1$ under the condition that $\sum_{\alpha}p_{\alpha} = 1$. This must be the expression for $\rho_A$ by uniqueness. We are able to construct the reduced density matrix for any state $\rho$ on $\hat{\mathcal{H}}$ using a von Neumann algebra on it. In the context of LQG, the natural von Neumann algebra is the algebra of gauge invariant observables in the holonomy-flux algebra. 

This expression \eqref{eq:Reduced_State} is qualitatively different than considering the state in $\mathcal{A}$ by itself. The difference is due to the presence of edge modes \cite{lin2018definingentropy}. Crucially, working solely within the algebra $\mathcal{A}$ leaves the structure of its center ambiguous, which obscures the boundary degrees of freedom captured by the edge modes. Our QEC-inspired approach remedies this by embedding the gauge-invariant algebra $\mathcal{A}$ into a larger, non-gauge-invariant 'ambient' algebra $\mathcal{M}$. This embedding fixes the center and naturally accounts for the edge modes. By starting with the algebra of non-gauge invariant operators and embedding the algebra of gauge invariant operators, the center is fixed and the edge modes appear naturally. From the ambient algebra, $\dim(\hat{\mathcal{H}}^{\text{edge}}_{\alpha})$ is the dimension of the space on which $\mathcal{A}$ is trivial on each sector. In this sense, the edge modes arise naturally from the center \cite{lin2018definingentropy}.

We now have all of the ingredients to construct the entropy of a state $\rho$ on a Hilbert space $\hat{\mathcal{H}}$ by choosing a von Neumann algebra $\mathcal{A}$:
\begin{equation}
    S(\rho,\mathcal{A}) = -\Tr(\rho_A\ln\rho_A)
\end{equation}
\begin{equation}
    S(\rho,\mathcal{A}) = \sum_{\alpha}-p_{\alpha}\ln p_{\alpha}+p_{\alpha}S(\rho_{A_{\alpha}})+\sum_{\alpha}p_{\alpha}\ln\dim(\hat{\mathcal{H}}^{\text{edge}}_{\alpha})
    \label{eq:Entropy}
\end{equation}
The presence of the Ryu-Takayanagi term is due directly to the edge modes. This is emblematic of the entanglement across the sectors. The renormalized trace can also be used to conveniently repackage the edge mode contribution into the normalization of the trace by requiring $\hat{\tr}_{\alpha}\mathbbm{1}_{\alpha} = 1$. This step makes it clear that the RT-like term is fundamentally a consequence of properly normalizing the algebra of observables for a subregion. The choice of the ambient algebra is what fixes this normalization unambiguously. This yields the renormalized density matrix:
\begin{equation}
    \hat{\rho}_A = \bigoplus_{\alpha} \bigl( p_{\alpha}\hat{\rho}_{A_{\alpha}}\otimes \mathbbm{1}_{\partial A_{\alpha}}\bigr)
    \label{eq:Renormalized_State}
\end{equation}
With the renormalized trace and state given by:
\begin{equation}
    \hat{\Tr} = \bigoplus_{\alpha}\frac{1}{\dim(\hat{\mathcal{H}}^{\text{edge}}_{\alpha})}\tr_{\alpha}
\end{equation}
\begin{equation}
    \hat{\rho}_{A_{\alpha}} = \dim(\hat{\mathcal{H}}^{\text{edge}}_{\alpha})\rho_{A_{\alpha}}
\end{equation}
Which leads to a simplified but equivalent entropy expression:
\begin{equation}
    S(\rho,\mathcal{A}) = \sum_{\alpha}p_{\alpha}\ln p_{\alpha}+p_{\alpha}S(\hat{\rho}_{A_{\alpha}})
\end{equation}
All of the edge mode contribution is now packaged into the normalization of the density matrix on each sector.

\section{Black Hole Entropy Area Law}\label{sec:Entropy_Law}
Black hole entropy has been calculated numerous ways within LQG \cite{perez2017bhloop}. The method
we discuss allows us to compute the entanglement entropy across any surface once $p_{\alpha}$
and the microscopic degrees of freedom in $\hat{\rho}_{A_{\alpha}}$ are specified. For the special case of a black
hole, we treat the system thermodynamically as a canonical ensemble. To illustrate the
method and elucidate important details, we first consider the entropy of a single edge
puncturing a surface.

\subsection{Single Edge Toy Model}
To mimic the canonical ensemble, we extremize the entropy of a single puncture while fixing the average area:
\begin{equation}
    a = \hat{\Tr}(\hat{\rho}_A\hat{A})
\end{equation}
Since the area operator is an element of the center, it is diagonalized by the reduced state:
\begin{equation}
    \hat{A} = 8\pi\gamma\ell_p^2\bigoplus_j \sqrt{j(j+1)}(\mathbbm{1}_j\otimes\mathbbm{1}_j)
\end{equation}
Applying this operator to our reduced state \eqref{eq:Renormalized_State}, we obtain a simple expression for the average area:
\begin{equation}
    a = 8\pi\gamma\ell_p^2\sum_jp_j\sqrt{j(j+1)}
\end{equation}
This leads to the variational principle that needs to be extremized:
\begin{equation}
    \delta\Bigl(\sum_j -p_j\ln p_j+p_jS(\hat{\rho}_{A_j})+2\pi\lambda\bigl( \frac{a}{8\pi\gamma\ell_p^2}-\sum_jp_j\sqrt{j(j+1)}\bigr)+\zeta\bigl(1-\sum_jp_j\bigr)\Bigr) = 0
    \label{eq:variational}
\end{equation}
The $2\pi\lambda$ is a convenient normalization. The probability distribution for the spin on the puncture follows:
\begin{equation}
    p_j = e^{-1-\zeta}e^{S(\hat{\rho}_{A_j})}e^{-2\pi\lambda\sqrt{j(j+1)}}
    \label{eq:distribution}
\end{equation}
It is convenient to define $Z = e^{\zeta+1}$ which fixes the normalization of the distribution:
\begin{equation}
    Z = \sum_j e^{S(\hat{\rho}_{A_j})}e^{-2\pi\lambda\sqrt{j(j+1)}}
\end{equation}
If we take the second variation, we find that it is positive. This implies that \eqref{eq:distribution} is a maximal entropy configuration. This probability yields the expected area dependence in the entropy:
\begin{equation}
    S(\hat{\rho}_A,\mathcal{A}) = \frac{\lambda}{\gamma}\frac{a}{4\ell_p^2}+\ln Z
\end{equation}
However, real surfaces do not contain just a single puncture. To capture multiple punctures,
we need to generalize our discussion to spin-network states on nontrivial graphs.

\subsection{Hilbert Space on a Graph}
Consider a closed and nontrivial graph in a three-dimensional compact space $\Sigma$. The graph contains a finite number of oriented edges $\mathfrak{e}$ and vertices $\mathfrak{v}$. Each edge is associated with an $L^2(SU(2))$ Hilbert space, so the quantum states on the whole graph belong to:
\begin{equation}
    \mathcal{H}_{\Gamma} := \bigotimes_{\mathfrak{e}}L^2(SU(2)) \cong \bigoplus_{\{j_{\mathfrak{e}}\}}\bigotimes_{
    \mathfrak{v}\in\Gamma} \bigl[\bigotimes_{\mathfrak{e},\mathfrak{v} = s(\mathfrak{e})}V_{j_{\mathfrak{e}}}\otimes \bigotimes_{\mathfrak{e},\mathfrak{v}=t(\mathfrak{e})}V^*_{j_{\mathfrak{e}}}\bigr]
\end{equation}
For a directed edge, we associate $V_j$ with the source vertex and $V^*_j$ with the target vertex. The $SU(2)$ gauge invariance projects each node Hilbert space to the invariant subspace:
\begin{equation}
    \widetilde{\mathcal{H}}_{\mathfrak{v}}(\{j_{\mathfrak{e}}\}) = \text{Inv}_{SU(2)}\bigl(\bigotimes_{\mathfrak{e},\mathfrak{v} = s(\mathfrak{e})}V_{j_{\mathfrak{e}}}\otimes \bigotimes_{\mathfrak{e},\mathfrak{v}=t(\mathfrak{e})}V^*_{j_{\mathfrak{e}}} \bigr)
\end{equation}
The gauge invariant node Hilbert space is the space of $SU(2)$ intertwiners. This Hilbert space has two special cases. It is $1$-dimensional if all of the spins are trivial and $0$-dimensional if the spins do not satisfy the triangle inequality. Enforcing gauge invariance at each vertex is what breaks the factorization and necessitates the need of von Neumann algebras. Therefore, the Hilbert space of all gauge invariant quantum states on the whole graph is:
\begin{equation}
    \widetilde{\mathcal{H}}_{\Gamma} = \bigoplus_{\{j_{\mathfrak{e}}\neq 0\} }\bigotimes_{\mathfrak{v}\in\Gamma}\widetilde{\mathcal{H}}_{\mathfrak{v}}(\{j_{\mathfrak{e}}\})
\end{equation}
This Hilbert space is spanned by linear combinations of spin-network states spanned by $(\Gamma,\{j_{\mathfrak{e}}\},\{i_{\mathfrak{v}}\})$. To obtain the full Hilbert space, we sum over all graphs:
\begin{equation}
    \mathcal{H}_{LQG} = \bigoplus_{\Gamma}\widetilde{\mathcal{H}}_{\Gamma}
\end{equation}
In order for the sum over all graphs to be well-defined, we need to impose cylindrical consistency to ensure that physically equivalent states defined on different graphs are not overcounted. Suppose that we have two graphs $\Gamma$ and $\Gamma'$ with two states $|\psi_{\Gamma}\rangle\in \widetilde{\mathcal{H}}_{\Gamma}$ and $|\psi_{\Gamma'}\rangle\in \widetilde{\mathcal{H}}_{\Gamma'}$. These two states are equivalent, or cylidrically consistent, if and only if:
\begin{equation}
    \iota_{\Gamma\rightarrow \Gamma''}|\psi_{\Gamma}\rangle = \iota_{\Gamma'\rightarrow\Gamma''}|\psi_{\Gamma'}\rangle
\end{equation}
where $\Gamma''$ is any common refinement of both $\Gamma$ and $\Gamma'$ and $\iota$ is an isometric embedding from a graph to its refinement. Pragmatically, the cylindrical consistency ensures that states that only differ by trivial edges on refinements are identified as the same. Recall the single edge Hilbert space of section \ref{sec:Single_Edge}. There, we established that the QEC construction can be understood through the cylindrical consistency of the graph refinement from a single-edge to two-edges connected by a bivalent vertex. We can generalize this perspective. We can understand the embedding $\iota_{\Gamma\rightarrow\Gamma''}$ as the encoding map from the logical Hilbert space $\widetilde{\mathcal{H}}_{\Gamma}$ to the physical Hilbert space $\widetilde{\mathcal{H}}_{\Gamma''}$. As a result, the QEC structures naturally lift to the full LQG Hilbert space. 

With the full Hilbert space, we can begin capturing surfaces. Begin by dividing the spatial slice into two regions $A$ and $\bar{A}$. The interface between $A$ and $\bar{A}$, $\mathfrak{S}$, is the surface that we consider the entanglement across. This surface is closed. Consider a graph $\Gamma$ that intersects $\mathfrak{S}$ with a number of edges and no vertices. We denote the edges of $\Gamma$ that intersect $\mathfrak{S}$ by $\mathfrak{e}_0$. We denote $\mathcal{J}$ as the collection of spins on the intersected edges, $\mathcal{J} = \{j_{\mathfrak{e}_0}\}$. If we fix $\{j_{\mathfrak{e}}\}$ on all edges, then $\{j_{\mathfrak{e}}\}$ also determines $\mathcal{J}$. Since $\mathfrak{S}$ contains no nodes, we can factorize $\bigotimes_{\mathfrak{v}}\widetilde{\mathcal{H}}_{\mathfrak{v}}\subset \widetilde{\mathcal{H}}_{\Gamma}$ over the subregions:
\begin{equation}
    \bigotimes_{\mathfrak{v}}\widetilde{\mathcal{H}}_{\mathfrak{v}} = \widetilde{\mathcal{H}}_A(\mathcal{J},\{j_{\mathfrak{e}}\}_{\mathfrak{e}\subset A})\otimes \widetilde{\mathcal{H}}_{\bar{A}}(\mathcal{J},\{j_{\mathfrak{e}}\}_{\mathfrak{e}\subset \bar{A}})
\end{equation}
Each of these new Hilbert spaces is a tensor product of all node Hilbert spaces in $A$ and $\bar{A}$ respectively. The factorization of the Hilbert space degrees of freedom can only be achieved if $\{j_{\mathfrak{e}_0}\}$ is fixed. In light of this factorization, the direct sum over spins of the graph Hilbert space can be viewed as a sum over edges inside $A$ and $\bar{A}$, with only those in $\mathcal{J}$ remaining:
\begin{equation}
    \widetilde{\mathcal{H}}_{\Gamma} = \bigoplus_{\mathcal{J}}\widetilde{\mathcal{H}}_{A,\mathcal{J}}\otimes \widetilde{\mathcal{H}}_{\bar{A},\mathcal{J}}
    \label{eq:LQG_Hilbert}
\end{equation}
Where the new Hilbert spaces capture the sum over edges that lie entirely with $A$ and $\bar{A}$ respectively. Fixing any $\mathcal{J}$, $\widetilde{\mathcal{H}}_A(\mathcal{J},\{j_{\mathfrak{e}}\}_{\mathfrak{e}\subset A})$ is nonzero if the triangle inequality is satisfied for all vertices in $A$. Now, to obtain the full Hilbert space, we sum over all possible punctures:
\begin{equation}
    \mathcal{H} = \bigoplus_{(\mathcal{P},\mathcal{J})}\widetilde{\mathcal{H}}_{A,(\mathcal{P},\mathcal{J})}\otimes \widetilde{\mathcal{H}}_{\bar{A},(\mathcal{P},\mathcal{J})}
    \label{eq:Puncture_Hilbert}
\end{equation}

The form of this Hilbert space is similar to that of a single edge of section $3$. The presence of the direct sum leads to the absence of the partial traces as we have previously discussed. Fortunately, letting $(\mathcal{P},\mathcal{J}) = \alpha$ be an abstract quantum number, we can easily use the von Neumann algebra techniques to generalize our single edge entropy to a whole surface.

\subsection{Entanglement Entropy Across a Surface}
The effective and renormalized reduced states on $\mathcal{H}$, given in \eqref{eq:LQG_Hilbert}, follow from the general construction of \eqref{eq:Renormalized_State}:
\begin{equation}
    \hat{\rho}_A = \bigoplus_{(\mathcal{P},\mathcal{J})}\bigl(p_{(\mathcal{P},\mathcal{J})}\hat{\rho}_{A,(\mathcal{P},\mathcal{J})}\otimes \mathbbm{1}_{\partial A,(\mathcal{P},\mathcal{J})}\bigr)
\end{equation}
For notational ease, we use $\alpha = (\mathcal{P},\mathcal{J})$ to represent the quantum numbers of the set of punctures and collection of spins on those punctures. Next, we need the von Neumann algebra in which to compute the entropy. This is done by constructing a von Neumann algebra out of the flux and holonomy operators. Following identically from \eqref{eq:Entropy}, the entropy of this reduced state is given by:
\begin{equation}
    S(\rho_A,\mathcal{A}) = \sum_{\alpha}-p_{\alpha}\ln p_{\alpha}+p_{\alpha}S(\rho_{A_{\alpha}})+\sum_{\alpha}p_{\alpha}\ln(2j_{\alpha}+1)
\end{equation}
The interpretation of the term is similar to the single edge toy model; it captures the geometric degrees of freedom entangled across the surface. The punctures in the surface naturally arise as the edge modes. Locally, the Hilbert space of each puncture schematically takes the form $\mathcal{H}_{\mathfrak{e}}\sim\bigoplus_j \bigl(V_j^A\otimes V^{*\bar{A}}_j\bigr)$. Since the $\dim(\hat{\mathcal{H}}^{\text{edge}}_{\alpha})$ captures the trivial action of the algebra on each edge, it picks out $\dim(V^{*\bar{A}}_j) = 2j+1$ for each edge. This leads to $\dim(\hat{\mathcal{H}}^{\text{edge}}_{\alpha}) = \prod_{\mathfrak{e}_0}(2j_{\mathfrak{e}_0}+1)$. The RT term naturally captures how the boundary geometry is encoded in the entanglement entropy.

In order to fix the average area for the ensemble, we first need to generalize the area operator to capture multiple punctures. For a single edge, the operator is given by $\hat{A}_{\mathfrak{e}_0} = 8\pi\gamma\ell_p^2\sqrt{\hat{J}_{\mathfrak{e}_0}\cdot\hat{J}_{\mathfrak{e}_0}}$. For multiple punctures with a collection of spins, we need to sum over all configurations:
\begin{equation}
    \hat{A} = \bigoplus_{(\mathcal{P},\mathcal{J})}8\pi\gamma\ell_p^2\sqrt{\hat{J}_{\mathfrak{e}_0}\cdot\hat{J}_{\mathfrak{e}_0}}
\end{equation}
The area operator resides in the center of our von Neumann algebra. Therefore, our reduced density matrix diagonalizes the area operator which simplifies the computation of the area of the state:
\begin{equation}
    A = \hat{\Tr}(\hat{\rho}_A\hat{A}) = 8\pi\gamma\ell_p^2\sum_{\alpha}p_{\alpha}\sqrt{j_{\alpha}(j_{\alpha}+1)}
\end{equation}
This leads to the variational principle being almost unchanged from the single puncture case in \eqref{eq:variational}. As a result, the entropy coming from the variational principle satisfies the expected area law:
\begin{equation}
    S(\rho_A,\mathcal{A}) = \frac{\lambda}{\gamma}\frac{A}{4\ell_p^2}+\ln \mathcal{Z}
\end{equation}
Where the new partition function is given by:
\begin{equation}
    \mathcal{Z} = \sum_{\alpha}(2j_{\alpha}+1)e^{S(\rho_{A_{\alpha}})}e^{-2\pi\lambda\sqrt{j_{\alpha}(j_{\alpha}+1)}}
\end{equation}

Our method reproduces the Bekenstein-Hawking entropy area law with an added term. However, the presence of $\lambda$ and the Immirzi parameter $\gamma$ makes our expression qualitatively different. This should be expected since the Bekenstein-Hawking entropy was computed under a semiclassical approximation where the spacetime is still smooth. Additionally, since we are deep in the Planckian regime, we would need to make contact with the low energy effective theory through renormalization \cite{ghosh2012scaling}. The renormalization of the prefactor can be equivalently treated as a renormalization of Newton's constant:
\begin{equation}
    [\ell_p^2]_{IR} = \frac{\lambda}{\lambda_0}[\ell_p^2]_{UV}
\end{equation}
This can also be understood through the divergence of the bulk entropy. The bulk entropy of quantum fields diverge due to the vacuum entanglement of the short wavelength modes across the horizon. The leading order divergence to the bulk entropy is proportional to the area, which then corresponds to the renormalization of Newton's constant.

The leading order divergence of the entropy can also be understood at the algebraic level. Type $III$ factor von Neumann algebras contain no renormalizable states. As a result, the entropy of each state is divergent. It was shown in \cite{soni2023type1} that type $III$ factors can be approximated by a direct sum of type $I$ factors under a proper limit. This limit is satisfied by the semiclassical limit $\ell_p^2\rightarrow 0$, resulting in the same leading order divergence in the bulk entropy.

The entropy of a black hole has been computed in LQG before by other means like state counting and the microcanonical ensemble. To make contact with these calculations in \cite{ghosh2011isolatedhorizons}, we have to consider the differences between the canonical ensembles. The key difference is whether the number of punctures is fixed. In order to construct the partition function using state counting, the number of punctures must be fixed, giving a new macroscopic observable. In our method, we sum over all possible punctures to construct the Hilbert space in \eqref{eq:Puncture_Hilbert}. This leads to considering a variable number of punctures over the configurations. In order to fix the number of punctures in our partition function, recall that $\alpha$ is an abstract index that captures which puncture is being considered and the collection of spins that can reside on the puncture. By fixing the number of punctures to $N$, we can clearly express the sum of sums:
\begin{equation}
    \mathcal{Z}_N = \sum_{j_1}\dots\sum_{j_N}(2j_{\alpha}+1)e^{S(\rho_{A_{\alpha}})}e^{-2\pi\lambda\sqrt{j_{\alpha}(j_{\alpha}+1)}}
\end{equation}
This further reduces by taking each puncture to be independent: 
\begin{equation}
    \mathcal{Z}_N = \bigl(\sum_j (2j+1)e^{-2\pi\lambda\sqrt{j(j+1)}}\bigr)^N
\end{equation}
This reduction also impacts the structure of the Hilbert space. Fixing the number of punctures and taking them to be independent has the effect of factorizing the Hilbert space. 
\begin{equation}
    \mathcal{H} \rightarrow \mathcal{H}_N = \bigotimes^N_{\alpha=1}\bigoplus_{j_{\alpha}}V_{j_{\alpha}}\otimes V^*_{j_{\alpha}}
\end{equation}
From the partition function, we can compute the area $\frac{A}{8\pi\gamma\ell_p^2} = -\partial_{\lambda}\ln Z_N$ which can be related to the number of punctures:
\begin{equation}
    \frac{A}{8\pi\gamma\ell_p^2} = 2\pi N \frac{\sum_j(2j+1)e^{-2\pi\lambda\sqrt{j(j+1)}}\sqrt{j(j+1)}}{\sum_j(2j+1)e^{-2\pi\lambda\sqrt{j(j+1)}}}
\end{equation}
The sums can be interpreted as a characteristic area $\hat{a}(\lambda)$ for a given Lagrange multiplier $\lambda$:
\begin{equation}
    \hat{a}(\lambda) = 8\pi\gamma\ell_p^2\frac{\sum_j(2j+1)e^{-2\pi\lambda\sqrt{j(j+1)}}\sqrt{j(j+1)}}{\sum_j(2j+1)e^{-2\pi\lambda\sqrt{j(j+1)}}}
\end{equation}
\begin{equation}
    N = \frac{A}{2\pi\hat{a}(\lambda)}
\end{equation}

The Lagrange multiplier $\lambda$ can be removed in favor of the thermodynamic temperature. This is done by defining the energy of the horizon in the semiclassical limit using the isolated horizon formalism. The semiclassical energy near the horizon is given by $E= \frac{A}{8\pi\ell_0}$, where $\ell_0$ is a fiducial length. Following standard statistical mechanics, the temperature is then defined by taking the derivative of the entropy with respect to the energy.
\begin{equation}
    \frac{1}{T} = \frac{\partial S}{\partial E} = \frac{2\pi\lambda}{\gamma}\frac{\ell_0}{\ell_p^2}
\end{equation}
This temperature can be exchanged for $2\pi\lambda$\footnote{$2\pi\lambda$ is used instead of $\lambda$ to make the different normalizations match.}. After setting the temperature to the Unruh temperature, $T_U = \frac{\ell_p^2}{2\pi\ell_0}$, both the expression for the entropy $S(\rho,\mathcal{A}) = \frac{A}{4\ell_p^2}$ and number of punctures $N = \frac{A}{2\pi\hat{a}(\gamma)}$ are consistent with previous results \cite{ghosh2012scaling}. 

It has been conjectured in \cite{lin2018definingentropy} that holographic entropy calculations utilizing the Ryu-Takayanagi formula corresponds explicitly to the full entanglement entropy. Their perspective asserts that the RT term arises from explicitly considering the full algebraic center. Considering the full center allows us to capture the full entanglement across the surface without restriction. This suggests that our quantum error correcting codes approach captures the full entanglement entropy across the surface via purely algebraic methods.

\section{Conclusion}
Throughout this work, we developed an algebraic framework, inspired by quantum error correcting codes \cite{harlow2016rtqec}, to compute the entanglement entropy across a surface in LQG. The central step is to specify a von Neumann algebra of gauge invariant operators $\mathcal{A}$ as the code subspace in the algebra of non-gauge invariant operators $\mathcal{M}$. The ambient algebra has a center which influences the edge modes of $\mathcal{A}$ and allows for the decomposition of states into a bulk contribution and an edge mode contribution. This immediately yields an entropy of the form \eqref{eq:Entropy} with the RT term capturing the boundary degrees of freedom. Specializing this analysis to black holes, we treat the punctures as a canonical ensemble. Maximizing the entropy subject to a fixed average area yields a distribution of spins on the punctures whose resulting entropy reproduces the area law:
\begin{equation}
    S(\rho,\mathcal{A}) = \frac{\lambda}{\gamma}\frac{A}{4\ell_p^2}+\ln \mathcal{Z}
\end{equation}
The leading coefficient matches the Bekenstein-Hawking value by the renormalization of Newton's constant and the proper matching in the semiclassical limit \cite{ghosh2012scaling}. The subleading $\ln\mathcal{Z}$ term contains state-dependent information about the edge mode degeneracies and sector bulk entropies.

Conceptually, this analysis emphasizes that the RT term is not an add-on but follows from the choice of algebra. Once the center is fixed, the edge modes become inevitable \cite{lin2018definingentropy} and precisely account for the puncture factor $\prod_{\mathfrak{e}_0}(2j_{\mathfrak{e}_0}+1)$. The quantum error correction framework naturally selects the non-gauge invariant algebra of operators which removes ambiguities due to working solely in the von Neumann algebra of gauge invariant operators. 

At its core, this construction is incomplete because it is purely kinematical. The states that we considered only obeyed gauge invariance at the vertices. Dynamical states need to be diffeomorphism invariant and invariant under the Hamiltonian constraint. The dynamical considerations should not impact the area law, but offer a method to compute $\lambda$. This is analogous to thermodynamics where the thermodynamic temperature determines the shape of the distribution but cannot be captured without microscopic dynamics. The techniques of quantum error correcting codes should immediately generalize once the dynamical von Neumann algebra is determined.

\ack{I would like to thank Muxin Han for discussions and guidance throughout this project. I was supported by the Florida Atlantic University Presidential Fellowship during this work.}




\bibliographystyle{iopart-num}
\bibliography{biblio.bib}

\end{document}